# In search of biomarkers and the ideotype of barley tolerant to water scarcity

Paweł Krajewski[1], Piotr Kachlicki[1], Anna Piasecka[2,1], Maria Surma[1], Anetta Kuczyńska[1], Krzysztof Mikołajczak[1], Piotr Ogrodowicz[1], Aneta Sawikowska[3,1], Hanna Ćwiek-Kupczyńska[1], Maciej Stobiecki[2], Paweł Rodziewicz[4], Łukasz Marczak[2]

[1] Institute of Plant Genetics, Polish Academy of Sciences, Strzeszyńska 34, 60-479 Poznań, Poland

[2] Institute of Bioorganic Chemistry, Polish Academy of Sciences, Noskowskiego 12/14, 61–704 Poznań, Poland

[3] Department of Mathematical and Statistical Methods, Poznań University of Life Sciences, Wojska Polskiego 28, Poznań, Poland

[4] Plant Biochemistry Laboratory, Department of Plant Biology, University of Copenhagen, 40 Thorvaldsensvej, DK-1871 Frederiksberg C, Copenhagen, Denmark

Abstract:

In barley plants, water shortage causes many changes on the morphological, physiological and biochemical levels resulting in the reduction of grain yield. In the present study the results of various experiments on the response of the same barley recombinant inbred lines to water shortage, including phenotypic, proteomic and metabolomic traits were integrated. Obtained results suggest that by a multi-omic approach it is possible to indicate proteomic and metabolomic traits important for reaction of barley plants to reduced water availability. Analysis of regression of drought effect (DE) for grain weight per plant on DE of proteomic and metabolomic traits allowed us to suggest ideotype of barley plants tolerant to water shortage. It was shown that grain weight under drought was determined significantly by six proteins in leaves and five in roots, the function of which were connected with defence mechanisms, ion/electron transport, carbon (in leaves) and nitrogen (in roots) metabolism, and in leaves additionally by two proteins of unknown function. Out of numerous metabolites detected in roots only Aspartic and Glutamic acids and one metabolite of unknown function, were found to have significant influence on grain weight per plant. The role of these traits as biomarkers, and especially as suggested targets of ideotype breeding, has to be further studied. One of the direction to be followed is genetic co-localization of proteomic, metabolomic and phenotypic traits in the genetic and physical maps of barley genome that can describe putative functional associations between traits; this is the next step of our analysis that is in progress.

**Financing:** The work was supported by the European Regional Development Fund through the Innovative Economy Program for Poland 2007–2013, project POLAPGEN-BD no. WND-POIG.01.03.01-00-101/08.

**Data availability:** The data supporting this report are partially available in the data set "POLAPGEN-BD integrative analysis demo" in the PlantPhenoDB database at www.cropnet.pl/plantphenodb/.

# 1. Introduction

Barley (*Hordeum vulgare* L.) is an important self-pollinating cereal crop grown under different environmental conditions. This species is characterized by high phenotypic variability and adaptability to unfavorable conditions. Barley breeding for a long term was focused on improving its yield capacity.

Research and breeding experience led to distinguishing of traits that should characterize highly yielding genotypes. Donald (1968) formulated the term "model plant or ideotype". An ideotype is a "hypothetical plant described in terms of traits that are thought to enhance genetic yield potential" (Rasmusson 1987). It was assumed that the breeding progress would be easier if the breeders knew the ideotype of the species they want to improve. Such ideal plants were then described for several species, including barley (Rasmusson 1987, Oosterom and Acevedo 1992). Ideotype of the plant should take into account biological properties of the species, user requirements and suitability for agricultural production under specific environmental conditions. The set of traits for an ideal barley plant depends on spike characters (2- or 6-rowed) and on what purpose the grain will be used for (food, brewing, fodder). Rasmusson (1987) proposed ideotype of 6-rowed spring barley grown in the upper midwestern USA. He considered 14 traits related to culm, head, leaf, growth duration, vegetative biomass and harvest index. The author concluded that such traits as plant stature, plant height, grain characteristics, vegetative biomass, harvest index are indeed associated with the yield in barley, but this is not always consistent with breeding experience. The author suggested that ideotype breeding should be a supplement to traditional breeding, but not replace the traditional method, i.e., selection based on yield performance *per se*.

Oosterom and Acevedo (1992) described two ideotypes of barley plants adapted to cold and drought: (1) characteristic for spring barley landraces originating from Australia and Jordania and (2) characteristic for landraces originating from eastern and north-eastern Syria. The first ideotype is a combination of an erect growth habit, early heading, light plant colour in winter, good early growth vigor and ability to recover from cold damage. The second ideotype is characterized, among others, by prostrate growth habit, dark green colour in winter, high level of cold tolerance and medium early heading. Both ideotypes should cover the ground to reduce soil evaporation in spring. The authors determined the usefulness of these two plant ideotypes as well as earliness as possible criteria for indirect selection for yield in Mediterranean environments, i.e. under low temperatures in winter and terminal-drought stress. They reported that early heading was positively correlated with grain yield, especially in low-yielding environments and concluded that the first described ideotype should be proposed for selection of barley lines adapted to Mediterranean environments with mild winters. These studies showed that it is difficult to define the ideotype of the plant, in terms of visible properties, adapted to the target environment.

Due to climate change another plant's features that are associated with high yield in diversified environments are sought after. The development of a plant resilient to environmental changes should currently be one of the most important breeding goals. To develop cultivars of a greater yield stability, genomic and biotechnological tools can be helpful. The results of research on primary phenotypic traits ("visible") in connection with genetic, molecular, proteomic and metabolomic research allow finding association between the primary and secondary features, and finding biomarkers useful for indirect selection of plants with target traits. Currently, the concept of an ideal plant includes primary traits and secondary characteristics modified at such a level that the target primary trait (e.g. yield) is increased. Features are sought that allow determination of the plant's resistance to stress, for example drought, related to control of the level of $CO_2$ assimilation and the rate of synthesis of compounds related to maintaining homeostasis by monitoring stable course of various biochemical processes. Tao et al. (2017) proposed climate-resilient and high-yielding future barley ideotypes for Boreal and Mediterranean climate in terms of phenology, leaf area, photosynthesis, drought tolerance and grain development. For grain yield formation out of eight simulation models, considered by the authors, four (APSIM, CropSyst, MCWALA and WOFOST) indicated that in Boreal and Mediterranean climate high-yielding barley plants should have a larger grain number, grain size and harvest index.

Adverse environmental conditions can affect plant's metabolism (e.g., De Mezer et al., 2014; Filek et al., 2014; Piasecka et al., 2017). Alterations inside an organism, at the level of invisible biochemical processes, lead to changes (at the morphological level) which usually result in the limitation of growth and development of plants and, finally, in the reduction of yield. Changes at the biochemical levels specific for one or a group of compounds, or their synthesis *de novo* in a specific tissue of plants growing under stress conditions, e.g. drought, may be a biochemical biomarker that can be used in the indirect selection of plants (Ernst 1999). Fernandez et al. (2016) defined a biomarker (i.e. biological marker) as "a characteristic that is objectively measured or evaluated as a predictor of plant performance". The authors distinguished genotypic biomarkers based on DNA polymorphism (e.g. single-nucleotide polymorphism, SNP) and phenotypic biomarkers such as transcript levels, protein levels, metabolite levels, enzyme activities, and images in different wavelengths. Several authors indicated that metabolic markers can be used to estimate plant performance under abiotic stresses (e.g. Degenkolbe et al. 2013, Fraire-Velázquez and Balderas-Hermández 2013, Nam et al. 2015).

Although it is difficult to develop one barley plant ideotype for many different target environments, the study of plant material in different environments and on different platforms, both with respect to phenotypic traits and biochemical characteristics, may help to describe the ideal plant - well yielding and resistant to stresses. In this type of research the need to study both aboveground and underground parts of crop plants is also important, as roots, leaves and stems play very different roles in the physiological processes, as well as due to the diversity of biochemical pathways taking place in these organs.

In barley, the response of plants to water scarcity has been analysed in many studies (Diab et al., 2004; Mansour et al., 2014; Mikołajczak et al. 2017, Ogrodowicz et al., 2017; Talamè et al., 2004; Teulat et al., 1998, 2001, 2003; Tondelli et al., 2014; Wójcik-Jagła et al., 2013). Water shortage causes many changes on the morphological, physiological and biochemical levels resulting in the limitation of growth and development, and, as aconsequence, in the reduction of yield. These changes were described, among others, for populations of recombinant inbred lines (RILs) derived from the crosses between European and Syrian genotypes, which were grown under drought conditions (Surma, Krajewski 2014). Response of those lines to water scarcity was evaluated in terms of yield and yield-related traits (Mikołajczak et al., 2016, 2017, Ogrodowicz et al., 2017), changes in metabolome and proteome (De Mezer et al. 2014, Rodziewicz et al. 2014, Chmielewska et al. 2016, Swarcewicz et al. 2017, Piasecka et al., 2017, Rodziewicz et al. 2019) as well changes occurring on physiological and molecular levels (Daszkowska-Golec & Szarejko, 2013, Filek et al. 2014, Bandurska et al. 2017, Baczek-Kwinta et al. 2018, Gudys et al. 2018).

The aim of the present study was to integrate the results of various experiments on the response of barley plants to water shortage, including phenotypic, proteomic and metabolomic traits, and to find biomarkers for drought tolerance, as well as to describe ideotype of barley plant tolerant to drought in terms of proteomic and metabolomic traits.

## 2. Material and methods

*2.1. Plant material and assays*

Material for the studies consisted of 100 spring barley (*Hordeum vulgare* L.) RILs derived from the cross Maresi × Cam/B1/CI08887//CI05761. Maresi is a German semidwarf cultivar with the pedigree Cebeco-6801/GB-1605//HA-46459-68, Cam/B1/CI08887//CI05761 (hereafter referred to as Cam/B1) is a Syrian breeding line adapted to dry environments (Mikołajczak et al. 2016). RILs were derived by the single-seed descent technique (Golden 1941) up to $F_8$ generation. The response of lines to temporal drought stress was assessed in the greenhouse experiment under drought water availability reduced to pF 3.2, with pF 2.2 for control condition (pF is a logarithm of the pressure *p,* expressed in centimeters of water head, necessary for removal of water from soil capillaries). Drought stress was applied for 10 days beginning from the 3-leaf stage (BBCH 13). Soil type, fertilization and maintaining the water regime were as it was described in Mikołajczak et al. (2017). The design of the experiment is shown in Fig. 1. On the fifth (T1) and 10th (T2) day of the drought, the plants were collected for biochemical analyses. Two pots for each genotype with 10 plants in each pot in control and drought treatments were designed for observation of phenological stages and for phenotyping after harvest.

The assays performed included (Fig. 1):

- Proteomic profiling in leaves and roots using 2D electrophoresis with quantitative analysis (by Image Master 2D Platinum software) of the observed trait "relative spot volume", in 4 biological and 2 technical replications (Rodziewicz et al. 2019). Protein identification was done on MALDI-TOF spectrometer. Approximately 1500 2D gels were observed; data were pre-processed with an own inter-gel matching procedure in Genstat (VSN International 2017).
- Profiling of primary metabolites in leaves and roots using GC/MS system (6890 N gas chromatograph, Agilent, USA, and a GCT Premier mass spectrometer, Waters, USA, with Waters MassLynx software version 4.1), in 2 biological and 4 technical replications (Swarcewicz et al. 2017). The observed trait was "concentration relative to internal standard". Metabolite identification was done using TargetSearch software with Golm Metabolome Database.
- Profiling of secondary metabolites in leaves at two time points T1 and T2, targeted mainly at phenolic compounds, by quantitative UPLC/UV analysis, in 4 biological replications (Piasecka et al. 2017). Approx. 1600 chromatograms (x 2 wavelengths) were observed; an alignment and quantitation procedure was programmed in R. The trait was "metabolite concentration". Metabolite identification was done by high-resolution Orbitrap analysis and database referencing.
- Phenotyping after harvesting with measurements of 26 traits (Mikołajczak et al. 2017).

The summary of traits taken into account in this report is shown in Table 1.

Table 1. Traits observed in the experiment

| Group of traits | Group ID | Number of observed traits | Number of traits transformed to drought effects and included in the analysis |
|---|---|---|---|
| Proteins leaves | PL | 257 | 119 |
| Proteins roots | PR | 381 | 95 |
| Primary metabolites leaves | ML | 99 | 84 |
| Primary metabolites roots | MR | 99 | 70 |
| Secondary metabolites leaves T1 | MT1 | 135 | 98 |
| Secondary metabolites leaves T2 | MT2 | 135 | 98 |
| Phenotypes | PHG | 26 | 26 |
| **Total** | | **1132** | **590** |

Units:
- PL, PR - %vol for spot in 2D electrophoresis (fraction of total volume of spots),
- ML, MR - intensities normalized to internal standard, unitless,
- MT1, MT2 - area under the curve (AUC) in chromatogram,
- PHG – depending on the trait.

*2.2. Data analysis*

For phenotypic traits, mean values for RILs observed under control and drought conditions were transformed to relative drought effects (RDE) using the formula RDE = 100 (mean in drought - mean in control) / mean in control; for other traits, absolute drought effect (ADE) were computed as ADE = (mean in drought – mean in control).

Computations related to correlation analysis, regression analysis and multivariate analysis (biplot) were performed in Genstat (VSN International 2017). Correction of *P* values with respect to multiple hypothesis testing was done using function *p.adjust* in R by Benjamini-Hochberg method. Correlation networks were visualized using Cytoscape (Shannon et al. 2003).

**3. Results**

Mean relative drought effects were negative for most of the phenotypic traits (Fig. 2A). Positive mean effects were observed for phenological traits describing flag leaf stage and heading date, and for protein content. Drought effects for proteomic traits were on average zero, and for metabolomic traits were mostly positive (Fig. 2B).

*3.1. Correlation analysis*

The correlation network for DE of all analysed traits, with 112 edges (corresponding to significant correlations listed in Table S1) revealed stronger correlations within groups of variables than between groups (Fig. 3). Leaf proteins (PL) were relatively weakly correlated with root proteins (PR) (2 edges). Primary metabolites in leaves (ML) and roots (MR) were linked by a number of significant correlations (6 edges), and both groups were linked to leaf proteins (PL). Secondary metabolites in leaves measured at T1 and T2 were linked by a number of significant correlations. Phenotypic traits could be divided into two correlated clusters, one with very few correlations with omic traits (cluster A: e.g. total yield, number of tillers, grain traits in lateral spikes, straw traits, harvest index, number of seeds, protein content), and one with a number of links to PL, ML, MR and MT2 (cluster B: e.g. length of stem, length of spikes, grain traits of main spike)

Analysis of correlations between DE-s for phenotypic and biochemical traits revealed direct links of traits PHG[3, 9, 10, 14, 20] to a number of PL, ML, MR and MT2 traits, and of traits PHG[4, 21, 24, 25] to another MR, PR and MT1 traits (Table S2, Fig. 4). Drought effect for grain size measured as 1000-grain weight (PHG[1]) was associated positively with DE for primary metabolite in leaves ML[25] and negatively with secondary metabolites MT2[78, 84, 121]. DE for heading stage (PHG[18]) was negatively correlated with two secondary metabolites in leaves measured at T1 (MT1[50] and [71]), but positively with secondary metabolite (MT2[114]) in T2. Length of main stem (PHG[3]) was positively associated with primary metabolites in leaves ML[24, 54, 58, 83, 94], and negative with ML[53, 55, 56, 66, 95] and PL[13, 126, 128, 201]. Length of main spike (PHG[10]) was correlated with the same MLs and in the same direction as PHG[3], and additionally positively with ML[3] and negatively with PL[126]. Grain weight per main spike (PHG[7]) and number of spikelets per main spike (PHG[9]) were also negatively correlated with primary metabolite ML[55]. Length of lateral spike (PHG[14]) was positively associated with primary metabolites in roots MR [44, 45]. Fertility of main spike (PHG[20]) and harvest index (PHG[25]) was correlated with the same MRs, but negatively (the latter only with MR[44]) (Tab. S2).

Examples of significant correlations between drought effects for phenotypic and omic traits are shown in Fig. 5. DE for length of main stem (PHG[3]) were negatively correlated with DE for Rubisco in leaves (PL[126]); the scatterplot shows a possibly nonlinear relationship. DE for straw weight per plant (PHG[4]) was also negatively correlated with Fructose in leaves (ML[55]). One can see that the determination coefficients of the visualised relationships are not high.

*3.2. Regression analysis*

Analysis of regression of drought effects for phenotypes on effects for omic traits shows a relatively large determination of stem and main spike traits by ML and MR traits, and of straw traits by PL (Tab. 2). Detailed analysis of regression coefficients showed dependence ($R^2$ above 30%) of DE for 16 phenotype traits on DE for primary metabolites in leaves (ML) and 12 PHG on DE for proteins in leaves (PL). The strongest determination was revealed for length of main spike and straw weight per pot and per plant by ML - $R^2$ amounted to 72.12, 66.64 and 61.22%, respectively. Root proteins (PR) determined mainly traits of lateral spikes – length, fertility and grain weight, as well number of seeds per plant, for which determination coefficient was above 30%. In turn, primary metabolites in roots (MR) affected strongly ($R^2 > 40\%$) straw weight per plant, 1000-grain weight, length of main stem, protein content and harvest index. Secondary metabolites in leaves at the beginning of drought (T1) influenced mainly heading stage, number of grains per main spike and harvest index - $R^2$ equal to 46.43, 41.00 and 32.87%, respectively, whereas at the end of drought (T2) – 1000-grain weight, length of main stem and main spike, and fertility of lateral spikes, for which determination coefficients were between 30.77 and 43.72% (Table S3). For DE of grain weight per plant (PHG2) only influence of PL and MR was important - $R^2$ 32.17% and 24.88%, respectively.

Table 2. Regression of DE for phenotypic traits on DE for groups of omic traits (values of $R^2$ for optimum regression equations obtained by forward stepwise selection)

| | Phenotype | Group of explanatory variables | | | | | | Mean $R^2$ |
|---|---|---|---|---|---|---|---|---|
| | | PL | PR | ML | MR | MT1 | MT2 | |
| 1 | 1000-grain weight | 33,96 | 21,93 | 33,46 | 49,26 | 13,63 | 36,47 | 31,45 |
| 2 | Grain weight per plant | 32,17 | 15,63 | 0,00 | 24,88 | 4,20 | 3,93 | 13,47 |
| 3 | Length of main stem | 50,23 | 28,82 | 57,65 | 47,30 | 18,94 | 38,83 | 40,30 |
| 4 | Straw weight per plant | 47,35 | 18,42 | 61,22 | 51,34 | 23,79 | 0,00 | 33,69 |
| 5 | Number of tillers per plant | 48,50 | 13,72 | 41,42 | 39,29 | 23,05 | 5,14 | 28,52 |
| 6 | Number of productive tillers per plant | 23,97 | 24,72 | 35,77 | 28,49 | 0,00 | 15,06 | 21,34 |
| 7 | Grain weight per main spike | 27,07 | 19,37 | 40,61 | 0,00 | 24,43 | 10,74 | 20,37 |
| 8 | Number of grains per main spike | 13,89 | 15,06 | 30,55 | 0,00 | 41,00 | 4,33 | 17,47 |
| 9 | Number of spikelets per main spike | 27,64 | 10,71 | 39,63 | 10,14 | 11,71 | 10,76 | 18,43 |
| 10 | Length of main spike | 43,56 | 19,03 | 72,12 | 44,25 | 28,68 | 30,77 | 39,73 |
| 11 | Grain weight per lateral spike | 17,41 | 41,47 | - | 11,27 | 17,59 | 0,00 | 17,55 |
| 12 | Number of grains per lateral spike | 0,00 | 24,97 | 0,00 | - | 21,26 | 13,31 | 11,91 |
| 13 | Number of spikelets per lateral spike | 41,37 | 18,53 | 38,41 | 9,56 | 8,03 | 0,00 | 19,32 |
| 14 | Length of lateral spike | 24,25 | 33,43 | 51,65 | 32,26 | 18,90 | - | 32,10 |
| 15 | Three leaves stage | 31,27 | 12,04 | 28,98 | 10,13 | 7,51 | 5,66 | 15,93 |
| 16 | Tillering stage | 21,91 | 22,24 | 48,64 | 5,18 | 11,53 | 16,61 | 21,02 |
| 17 | Flag leaf stage | 54,34 | 33,78 | 27,76 | 12,90 | 9,98 | 10,16 | 24,82 |
| 18 | Heading stage | 20,62 | 17,16 | 19,60 | 0,00 | 46,43 | 17,32 | 20,19 |
| 19 | Productivity of tillers | 14,94 | 9,80 | 7,34 | 9,44 | 11,24 | 3,39 | 9,36 |
| 20 | Fertility of main spike | 38,18 | 5,48 | 40,12 | 21,64 | 20,10 | 8,67 | 22,36 |
| 21 | Fertility of lateral spike | 4,48 | 38,04 | 45,09 | 11,57 | 17,76 | 43,72 | 26,78 |
| 22 | Protein content in grain | 39,43 | 21,22 | 0,00 | 42,45 | 7,86 | 24,57 | 22,59 |
| 23 | Grain weight per pot | 18,23 | 9,58 | 0,00 | 0,00 | 0,00 | 8,29 | 6,02 |
| 24 | Straw weight per pot | 32,33 | 13,80 | 66,64 | 15,54 | 18,52 | 10,23 | 26,18 |
| 25 | Harvest index | 19,66 | 24,22 | 25,67 | 47,00 | 32,87 | 18,23 | 27,94 |
| 26 | Number of seeds per plant | 12,19 | 42,70 | 41,49 | 32,78 | 0,00 | 5,17 | 22,39 |

Some interpretation of the set of $R^2$ values given in Table 2 can be obtained by a biplot (Fig. 6). Firstly, the biplot shows a lower contribution of MT1 traits than of other groups of traits to the determination of DE for phenotypic traits. Secondly, a contrast along the second principal axis can be seen between PL and PR traits, with PL being highly associated with straw weight (PHG[4]) and number of tillers (PHG[5]), and PR being associated with fertility (PHG[21]) and number of seeds (PHG[26]). Thirdly, along the same axis, a contrast between fertility of main spike (PHG[20]), related to PL traits. and fertility of lateral spikes (PHG[21]), related to PR traits, can be observed.

Examples of relationships between observed and predicted drought effects for phenotypic traits are shown in Fig. 7.

*3.3. Ideotype*

The correlation sub-network of RDE for phenotypic traits correlated with RDE for grain weight per plant and directly correlated with ADE for biochemical traits are shown in Fig. 8. It can be noticed that none of the omics features correlates directly with the grain weight per plant, but only indirectly, mainly through 1000-grain weight (PL[191], ML[25], and MT2[78, 84, 121]), straw weight per plant (PR[117], MR[6], ML[55], MT1[20]), and grain weight per lateral spike (PR[208]).

A picture of the direct determination of DE for grain weight by DE for omic traits is given in Table 3. The sets of explanatory traits largely differ from the sets shown in Fig. 8. The best determination is obtained by using the PL and MR traits, however, the determination coefficients are still l not high.

Table 3. Best regression equations for DE of grain weight on DE of omic traits obtained in groups of variables

| Group | Variable ID | Name | Functional/structural group | Regression coefficient | $R^2$ |
|---|---|---|---|---|---|
| PL | PL[30] | ATP-dependent Clp protease ATP-binding subunit ClpC | Defence | 2,33 | 33.17 |
|  | PL[37] | Chloroplast inner envelope protein (putative) | Ion/electron transport | 2,32 |  |
|  | PL[59] | Fructose-1,6-bisphosphate aldolase | Carbon metabolism | -2,71 |  |
|  | PL[181] | UTP-glucose-1-phosphate uridylyltransferase | Carbon metabolism | 3,51 |  |
|  | PL[191] | Unknown | Unknown | -2,67 |  |
|  | PL[206] | Unknown | Unknown | -3,98 |  |
| PR | PR[50] | ATP synthase subunit beta, mitochondrial | Ion/Electron transport | 2,84 | 15.63 |
|  | PR[79] | Dehydroascorbate reductase | Defence | -2,61 |  |
|  | PR[133] | Heat shock protein 70 kDa | Defence | -2,30 |  |
|  | PR[153] | Methionine synthase | Nitrogen metabolism | -2,45 |  |
| MR | MR[31] | A152002-Aspartic acid (3TMS) | amino acids | 5,11 | 24.88 |
|  | MR[38] | A163001-Glutamic acid (3TMS) | amino acids | -2,60 |  |
|  | MR[96] | A311002-NA | unknown-other | -2,46 |  |
| MT1 | MT1[70] | Isovitexin/Isovitexin 2''-O-rhamnoside |  | 2,17 | 4.20 |
|  | MT2[102] | UPLC_UV_signal_078 |  | -2,22 | 3.93 |

Taking into account the regression coefficients and the direction of DE for proteins and metabolites given in Table 3 drought tolerant barley plants, i.e. with a low decrease of grain weight per plant under water shortage, should be characterized by :

- low decrease of 1000-grain weight, straw weight per plant and grain weight per lateral spike,

- low increase in drought amount of:
    ATP-dependent Clp protease ATP-binding subunit ClpC (PL[30]) in leaves,
    ATP synthase subunit beta, mitochondrial (PR[50]) and metabolite A152002-Aspartic acid (3TMS) (MR[31]) in roots,
- low decrease under drought amount of:
    UTP-glucose-1-phosphate uridylyltransferase (PL[181] and Chloroplast inner envelope protein (PL[37]) in leaves,
- high decrease in drought amount of:
    Fructose-1,6-bisphosphate aldolase (PL[59]) in leaves
    and Methionine synthase (PR[153]) in roots,
- high increase under drought amount of:
    Dehydroascorbate reductase (PR[79]) and Heat shock protein 70 kD (PR[133]), and
    metabolites: A163001-Glutamic acid (3TMS) (MR[38]) and A311002-NA (MR[96]) in roots.

Some insight into relative contribution into determination of DE for grain weight by DE for other observed traits can be gained by considering a difference between the process of forward selection of best regressors with different sets of potential ones (Table 4). In case A, phenotypic traits were taken as the full set of explanatory variables; in case B, the full set consisted of all observed traits. Until $5^{th}$ step the selection process was the same. Then, in case B, two PL traits were selected instead of further phenotypic traits - as it was in case A. This shows some advantage of observation of DE for proteomic traits over DE for phenotypic traits for prediction of drought effects for yield.

Table 4. Forward selection of regression equation of DE for grain weight on DE for phenotypic traits or on DE for phenotypic and omic traits

| A. Selection from phenotypic variables only | | | | B. Selection from phenotypic and biochemical variables | | |
|---|---|---|---|---|---|---|
| Explanatory variable | Name | Regression coefficient | s.e. | Explanatory variable | Regression coefficient | s.e. |
| PHG[26] | Number of seeds per plant | 0,53 | 0,05 | PHG[26] | 0,47 | 0,05 |
| PHG[1] | 1000-grain weight | 0,28 | 0,07 | PHG[1] | 0,34 | 0,06 |
| PHG[11] | Grain weight per lateral spike | 0,40 | 0,08 | PHG[11] | 0,25 | 0,05 |
| PHG[4] | Straw weight per plant | 0,16 | 0,04 | PHG[4] | 0,18 | 0,04 |
| PHG[24] | Straw weight per pot | -0,13 | 0,05 | PHG[24] | -0,13 | 0,05 |
| PHG[16] | Tillering stage | -0,81 | 0,35 | PL[206] (unknown) | -8,98 | 2,88 |
| PHG[12] | Number of grains per lateral spike | -0,20 | 0,10 | PL[30] (ATP-dependent Clp protease ATP-binding subunit ClpC) | 4,64 | 2,13 |
| PHG[10] | Length of main spike | 0,15 | 0,08 | | | |
| $R^2 = 80.7$ | | | | $R^2 = 81.4$ | | |

## 4. Discussion

In this report we present an attempt to integrate results of proteomic, metabolomic and phenotypic assays performed in an experiment studying reaction of a set of barley recombinant inbred lines to reduced water availability. Performing such multi-omic investigations is challenging for many reasons. Firstly, collection of materials for all measurements according to an experimental plan appropriate for sound statistical inference requires a large number of experimental units (pots) treated in the same way and kept in a uniform environment. Secondly, the studied pool of genotypes must be large to obtain meaningful relationships between measured traits. Thirdly, it requires the ability to perform a number of specialized biochemical measurements within short time ranges adapted to plant phenology, at a number of laboratories collaborating on the collection of plant materials. In our case, overcoming these challenges was possible due to experimental capacity, selection of a set of RILs and agreements and detailed planning achieved within POLAPGEN-BD project (www.polapgen.eu).

The set of traits analysed jointly in the presented analysis consisted of proteomic, metabolomic and phenotypic features measured by different protocols. With the number of studied genotypes being 100, and with the numbers of biological and technical replications being sufficient for the task, the numbers of measurements were very large. Special data pre-processing algorithms had to be devised, e.g. for the alignment of multiple gel data or multiple chromatograms. Subsequently, various data normalizations were performed. From the general methodological point of view it can be of interest how all these numerical procedures, in particular normalizations, affect the final analysis results (for example, the topology of correlation networks). We do not have a general answer to this question, but we think that all necessary care was taken in data analysis, in particular to statistically reduce the number of outliers, correct testing results for multiple hypothesis testing, and – in the statistical inference part – use standard procedures (like correlation or regression) with relatively simple interpretation. This was possible because in the analysis we really did not meet the "big data" challenges in the statistical sense, i.e., the situations in which the number of traits exceeds substantially the number of cases, although the number analyzed traits was large. Application of some more advanced methods like machine learning algorithms was also tried but did not provide additional insights.

The meaning of our data integration was dictated from the beginning by the choice to transform the data to drought effects, i.e., to study not the performance of the genotypes in different environments per se, but their reaction to the drought treatment. This directed us towards inducible biomarkers, not constitutive ones, as discussed by Fernandez et al. (2016). We assumed that not the absolute value of a protein or metabolite, but its (de)accumulation under drought, can explain the phenotypic reactions of plants. This type of analysis simplifies the considerations, however, probably cannot be used instead of a separate analysis of plants under control and drought conditions, which is also planned.

The correlation analysis indicated strong relationships of traits within considered groups; this, however, did not obscured the inter-group correlations which existence is important for our final aim. We obtained more significant correlations between leaf and root proteins than between leaf and root primary metabolites. Moreover, a subset of phenotypic traits could be distinguished that correlated with omic traits; it contained stem and spikes length measurements and grain traits of main spike. We found secondary metabolites with drought effects correlated both negatively (Isovitexin, Sinapic acid) and positively (unknown) with reaction of plants in terms of earliness of heading; this is interesting as early heading was named as important in the context of barley ideotype by Oosterom and Acevedo (1992). Drought effects for length of lateral spikes were positively associated with effects for some primary metabolites in roots (Glutamine, Putrecsine); this could be explained by the fact that lateral spikes are developed later than main ones, when the development of the root system is more decisive. The strength of observed correlations was at the level observed also by other authors, e.g., for quality traits by Heuberger et al. (2014).

The regression analysis supplemented the correlation networks by revealing relationships that may be weaker, but remain significant after elimination of stronger effects of other traits. The strongest determination coefficients were obtained for stem length and straw weight traits when considering primary leaf metabolites as explanatory traits. However, effects for primary metabolites in roots were also important for straw and stem traits. Regression analysis confirmed relatively strong determination of lateral spike characteristics by root proteins.

The statistical analysis of drought effects should be supplemented by functional analysis of the discovered significant effects. By using one of the tools for such analyses, MetaboAnalyst (Chong et al. 2019), we were able to indicate the pathways of Aminoacyl-tRNA biosynthesis, Alanine aspartate and Glutamate metabolism, and Arginine and Proline metabolism to be enriched in leaf primary metabolites with significant drought effects. However, drought effects of only a few metabolites causing this enrichment could be indicated as explanatory for phenotypic drought effects. This means that the effect of our analysis is identification of some inducible biomarkers, rather isolated in various parts of the whole metabolic system, but not of branches of the metabolism that are decisive for plant resistance to drought.

Taken together, the results of omic and phenotypic data integration suggest that by a mulit-omic approach it is possible to indicate proteomic and metabolomic traits important for reaction of barley plants to reduced water availability. Analysis of regression of DE for single plant grain weight on DE of proteomic and metabolomic traits allowed us to suggest ideotype of barley plants tolerant to water shortage. Grain weight drought effect was determined significantly by six proteins in leaves and five in roots which were connected with defence mechanisms, ion/electron transport, carbon (in leaves) and nitrogen (in roots) metabolism, as well in leaves by two other proteins of unknown function. Among numerous metabolites detected in roots only three, two amino acids (Aspartic and Glutamic) and one of unknown function, were found to have significant influence on grain weight per plant. The role of these traits as biomarkers, and especially as suggested targets of ideotype breeding, has to be further studied. One of the direction to be followed is genetic co-localization of proteomic, metabolomic and phenotypic traits in the genetic and physical maps of barley genome that can describe putative functional associations between traits; this is the next step of our analysis that is in progress.

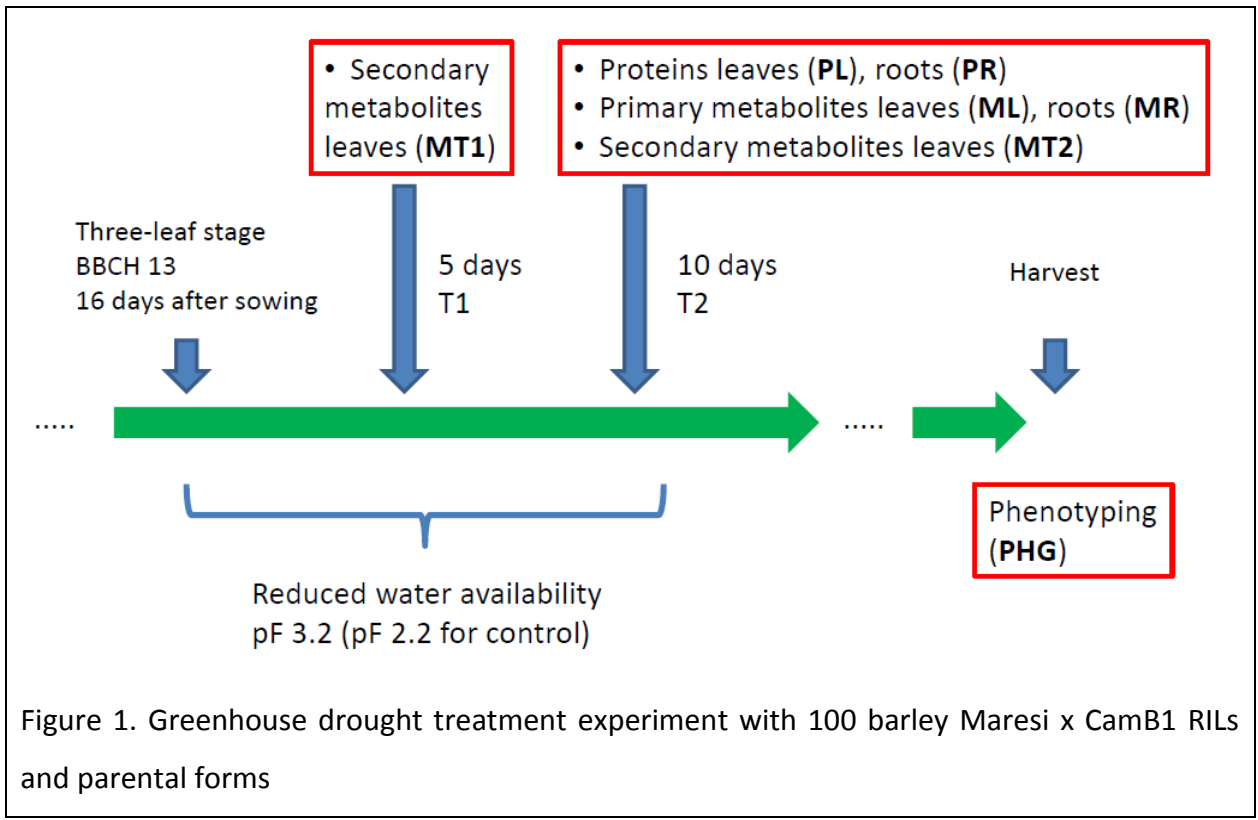

Figure 1. Greenhouse drought treatment experiment with 100 barley Maresi x CamB1 RILs and parental forms

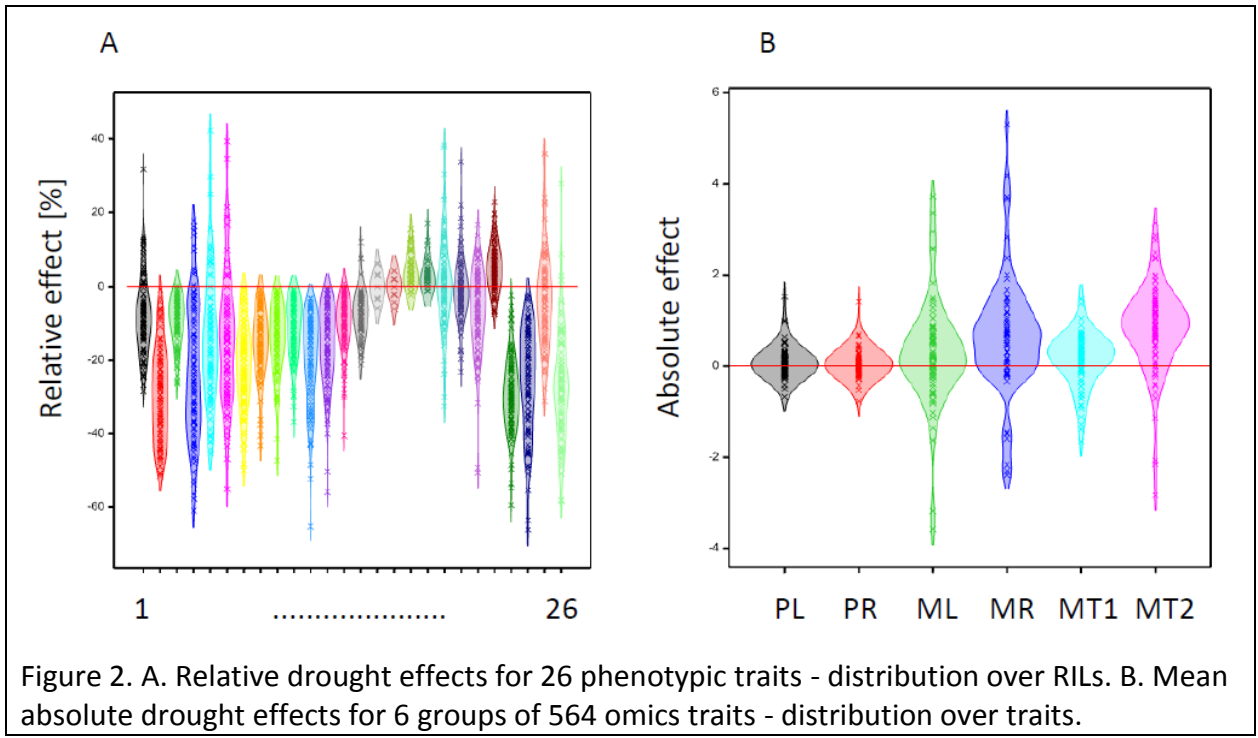

Figure 2. A. Relative drought effects for 26 phenotypic traits - distribution over RILs. B. Mean absolute drought effects for 6 groups of 564 omics traits - distribution over traits.

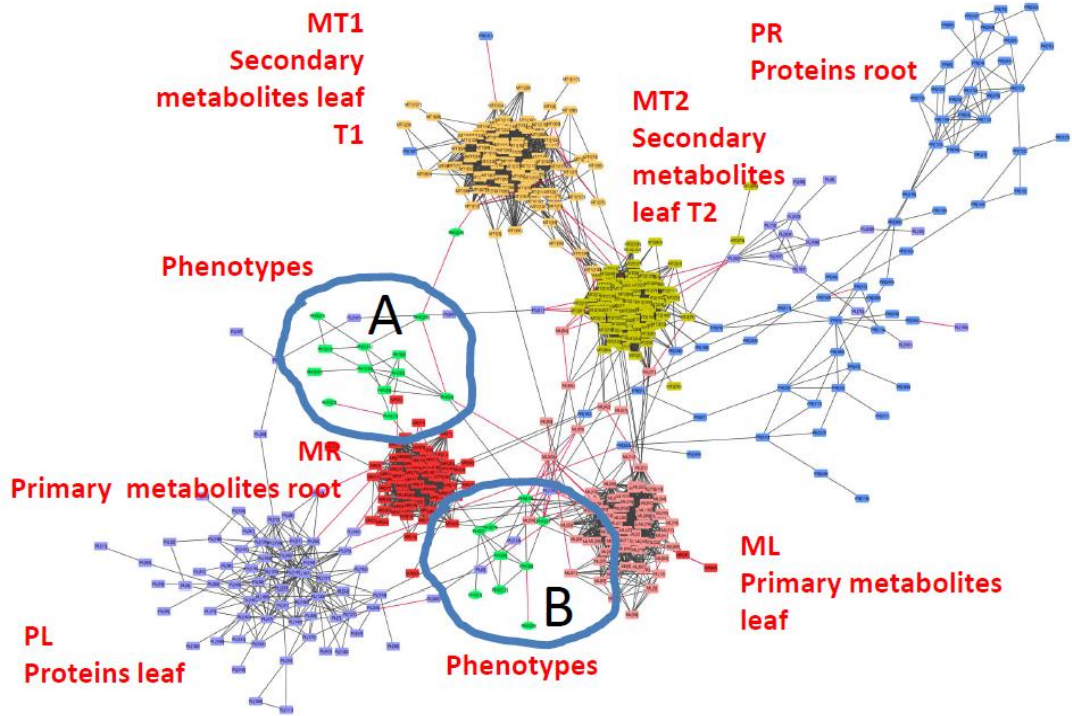

Figure 3. Correlation network for relative drought effects for phenotypic traits and absolute drought effects for omic traits. Edges drawn for 5854 correlations with Benjamini-Hochberg FDR < 0.001 (3.3% selected out of all 173755 correlations; filtering corresponded to P-value < 0.00003).

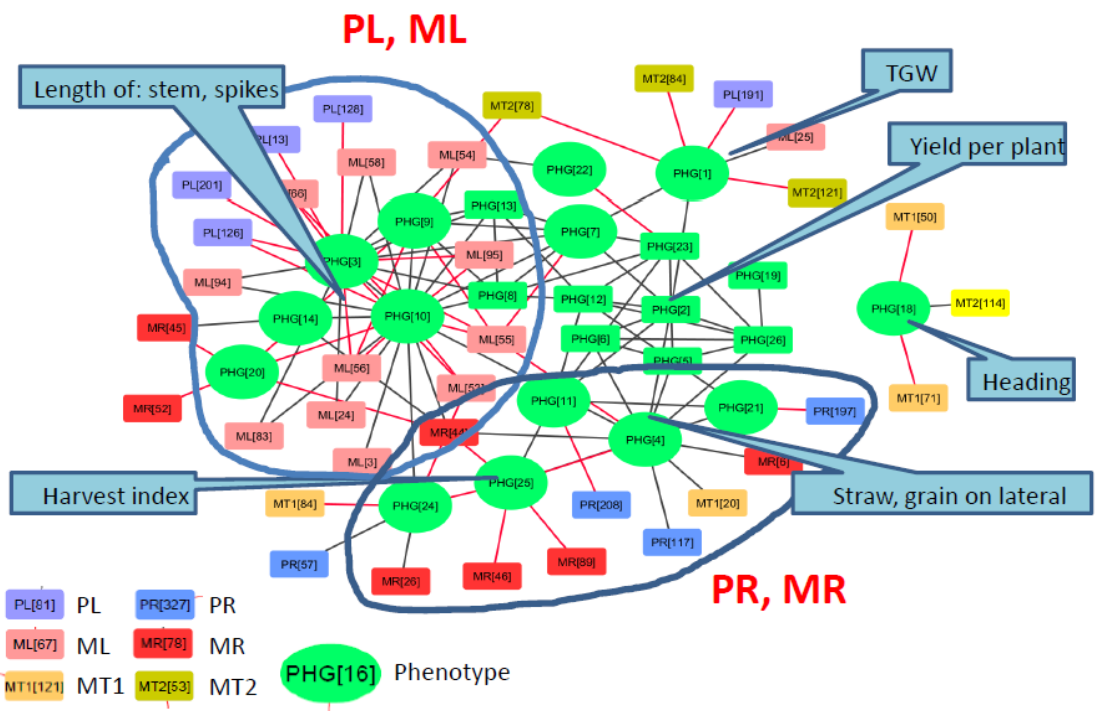

Figure 4. Correlation sub-network for phenotypes and their neighbours . Benjamini-Hochberg FDR < 0.05, 112 correlations out of 14664 (0.76%), corresponds to P-value < 0.0003.

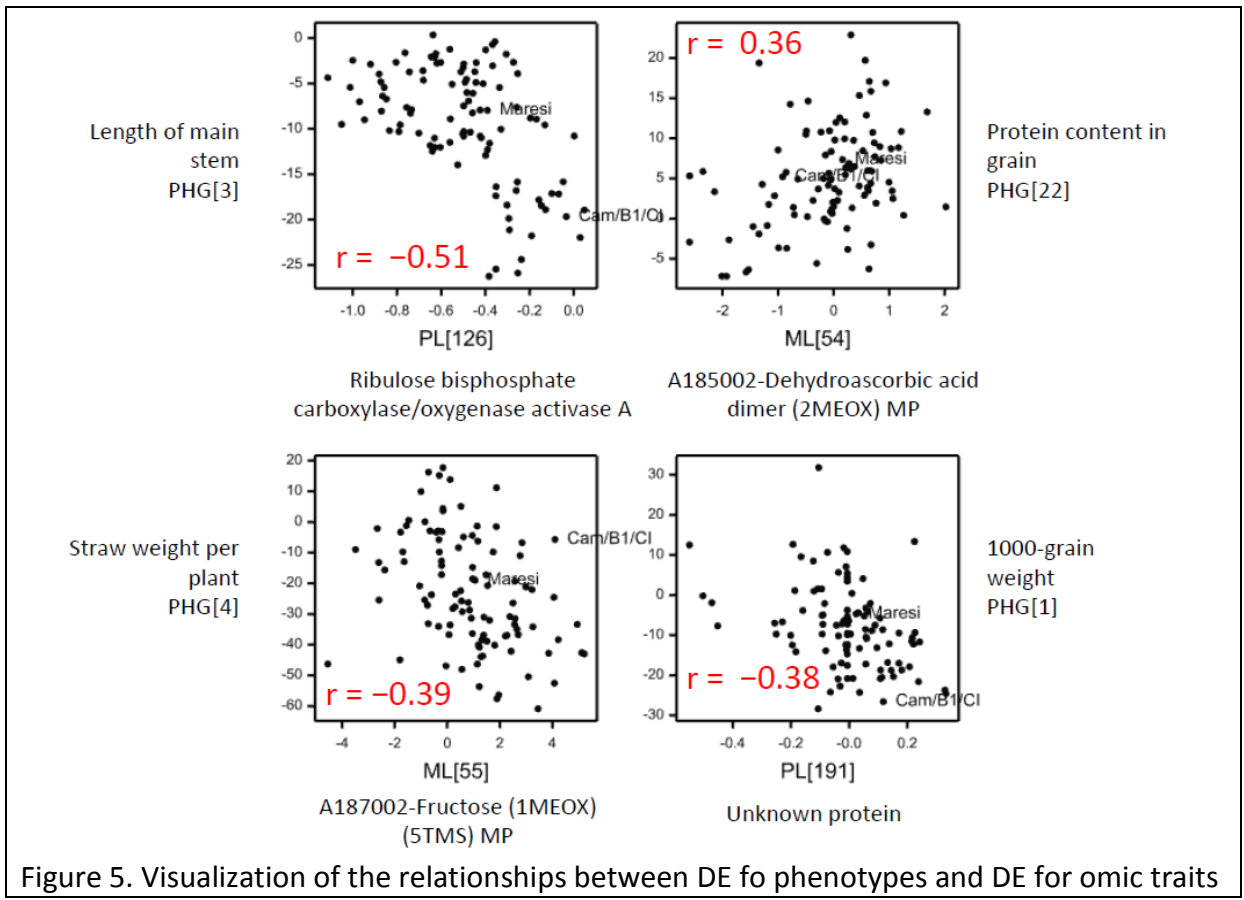

Figure 5. Visualization of the relationships between DE fo phenotypes and DE for omic traits

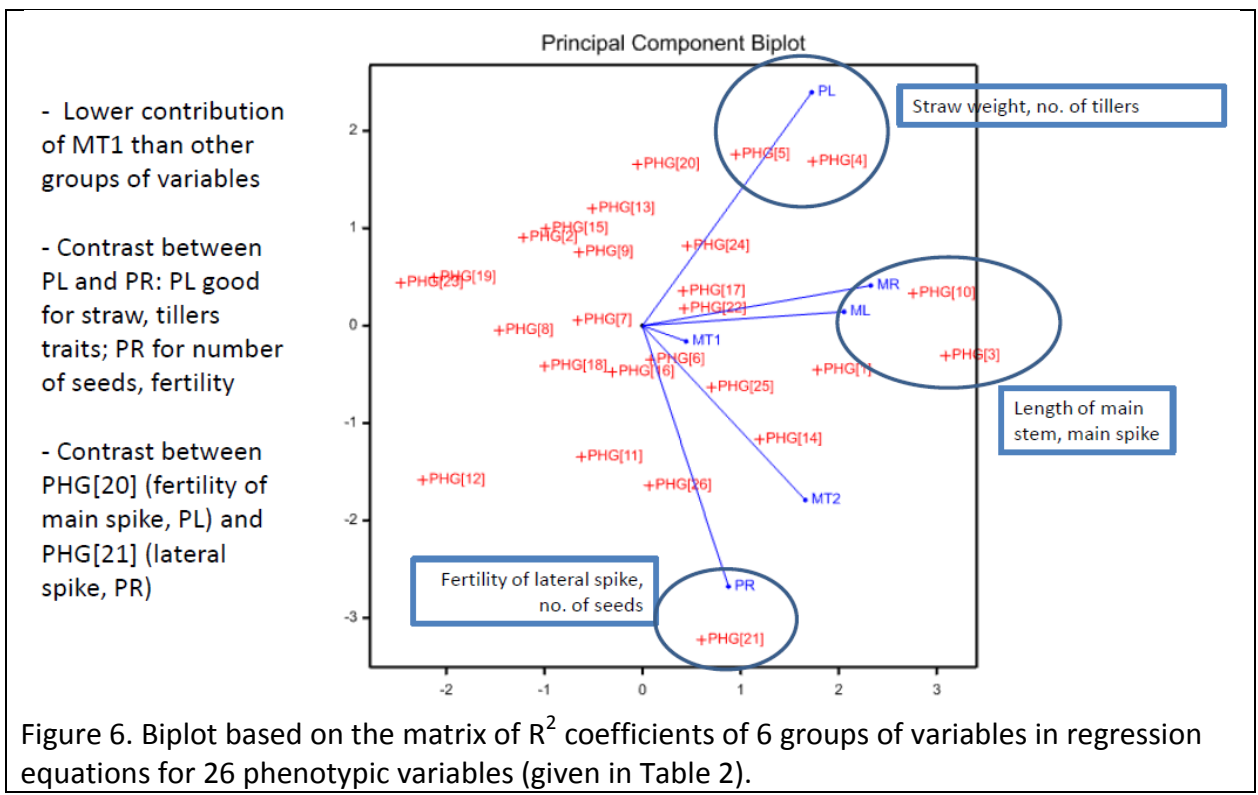

Figure 6. Biplot based on the matrix of $R^2$ coefficients of 6 groups of variables in regression equations for 26 phenotypic variables (given in Table 2).

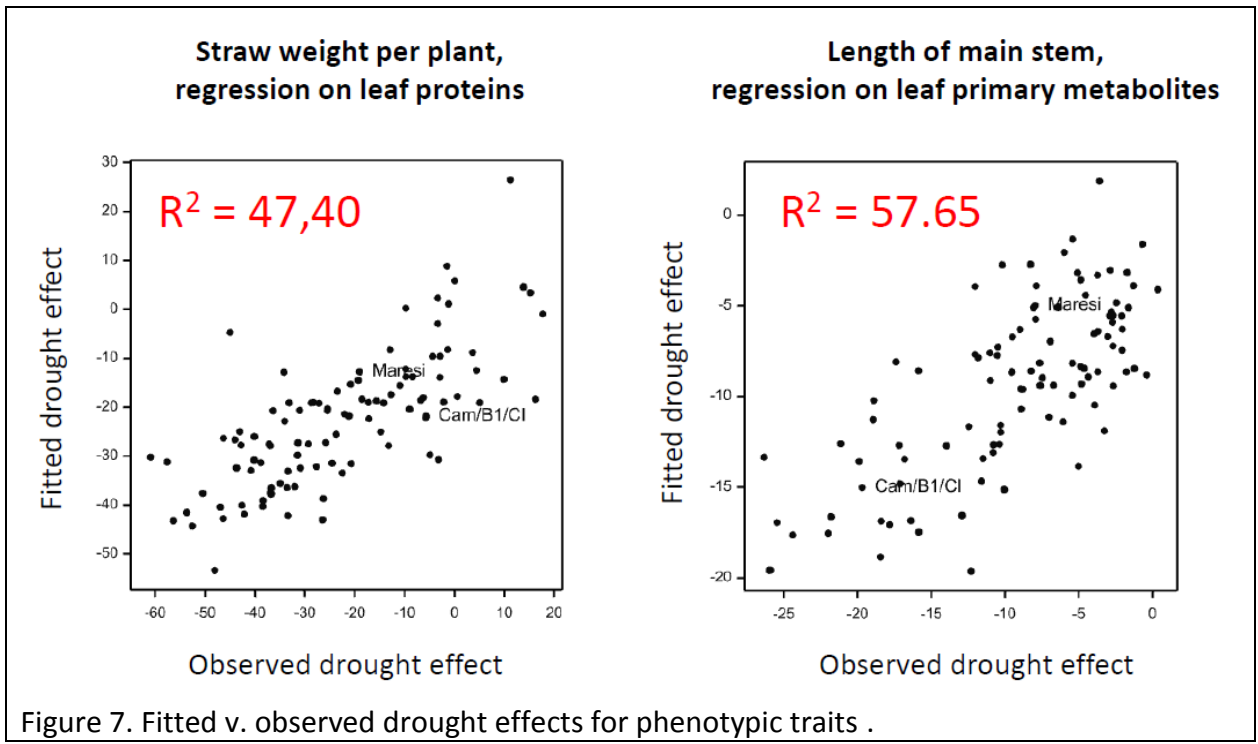

Figure 7. Fitted v. observed drought effects for phenotypic traits .

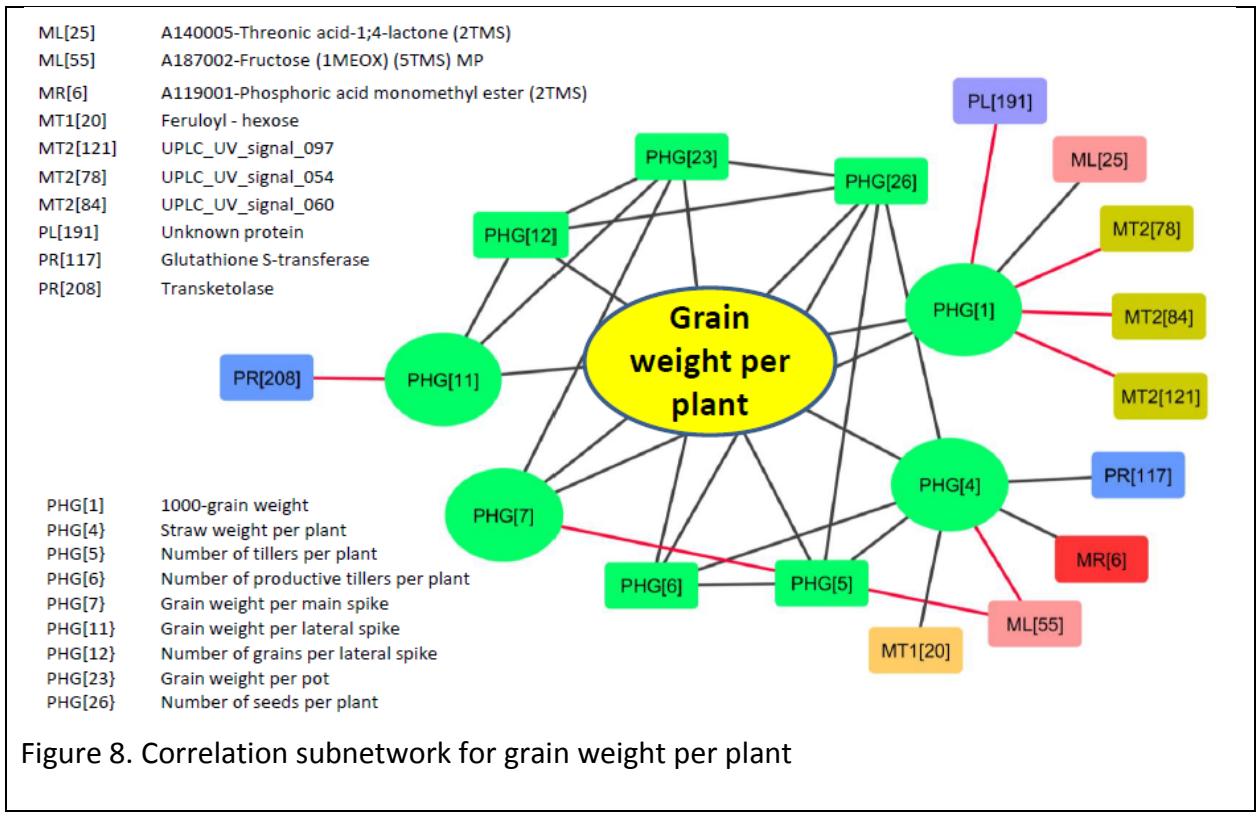

Figure 8. Correlation subnetwork for grain weight per plant

Table S1. Statistically significant correlations between omic traits (corresponding to edges in Figure 3)

| Group of traits | Trait ID | Trait name | Category | Group of traits | Trait ID | Trait name | Category | Correlation |
|---|---|---|---|---|---|---|---|---|
| PL | PL[5] | 20 kDa chaperonin, chloroplastic | Defence | ML | ML[15] | A132003-Proline (2TMS) | amino acids | -0,40 |
| PL | PL[29] | ATP-dependent Clp protease ATP-binding subunit ClpC | Defence | MR | MR[89] | A264001-Sucrose (8TMS) | carbohydrates | -0,41 |
| PL | PL[30] | ATP-dependent Clp protease ATP-binding subunit ClpC | Defence | MR | MR[15] | A132003-Proline (2TMS) | amino acids | -0,40 |
| PL | PL[32] | ATP-dependent zinc metalloprotease FTSH 10 | Defence | MT2 | MT2[3] | 5-Feruloylquinic acid | | -0,39 |
| PL | PL[32] | ATP-dependent zinc metalloprotease FTSH 10 | Defence | MT2 | MT2[17] | Chrysoeriol 7-O-rhamnosylglucoside/Tricin 7-O-rhamnosylglucoside/Tricin 7-O-arabinosylglucoside/Tricin 7-O-diglucoside/Tricin 7-O-glucoside | | -0,42 |
| PL | PL[32] | ATP-dependent zinc metalloprotease FTSH 10 | Defence | MT2 | MT2[110] | UPLC_UV_signal_086 | | -0,40 |
| PL | PL[41] | Cytochrome b6-f complex iron-sulfur subunit | Photosynthesis | MT2 | MT2[15] | Chrysoeriol 7-O-arabinosylglucoside | | -0,40 |
| PL | PL[41] | Cytochrome b6-f complex iron-sulfur subunit | Photosynthesis | MT2 | MT2[126] | UPLC_UV_signal_102 | | -0,42 |
| PL | PL[41] | Cytochrome b6-f complex iron-sulfur subunit | Photosynthesis | MT2 | MT2[132] | UPLC_UV_signal_108 | | -0,40 |
| PL | PL[78] | Heat shock 70 kDa protein | Defence | PR | PR[323] | Unknown protein | Unknown | -0,39 |
| PL | PL[108] | Phosphoglycerate kinase, chloroplastic | Photosynthesis | ML | ML[59] | A192003-Lysine (4TMS) | amino acids | 0,40 |
| PL | PL[126] | Ribulose bisphosphate carboxylase/oxygenase activase A | Photosynthesis | ML | ML[24] | A140001-Threonine (3TMS) | amino acids | -0,51 |
| PL | PL[126] | Ribulose bisphosphate carboxylase/oxygenase activase A | Photosynthesis | ML | ML[54] | A185002-Dehydroascorbic acid dimer (2MEOX) MP | sugar acids | -0,42 |
| PL | PL[126] | Ribulose bisphosphate carboxylase/oxygenase activase A | Photosynthesis | ML | ML[55] | A187002-Fructose (1MEOX) (5TMS) MP | carbohydrates | 0,55 |
| PL | PL[126] | Ribulose bisphosphate carboxylase/oxygenase activase A | Photosynthesis | ML | ML[66] | A203003-NA | unknown-other | 0,40 |
| PL | PL[126] | Ribulose bisphosphate carboxylase/oxygenase activase A | Photosynthesis | ML | ML[73] | A217007-NA | unknown-other | -0,38 |
| PL | PL[128] | Ribulose bisphosphate carboxylase/oxygenase activase A | Photosynthesis | ML | ML[53] | A185001-Quinic acid (5TMS) | other carboxylic acids | 0,40 |
| PL | PL[140] | Ribulose bisphosphate carboxylase/oxygenase activase B | Photosynthesis | MR | MR[20] | A137001-Fumaric acid (2TMS) | Krebs cycle acids | -0,42 |
| PL | PL[140] | Ribulose bisphosphate carboxylase/oxygenase activase B | Photosynthesis | MR | MR[21] | A137004-2-Piperidinecarboxylic acid (2TMS) | amino acids | -0,41 |
| PL | PL[192] | Unknown protein | Unknown | MR | MR[44] | A174008-Glutamine (4TMS) | amino acids | 0,42 |
| PL | PL[199] | Unknown protein | Unknown | PR | PR[184] | Proteasome subunit alpha type-2 | Defence | -0,46 |
| PR | PR[79] | Dehydroascorbate reductase | Defence | MT2 | MT2[31] | Isoorientin 2''-O-glucoside sinapide/Isovitexin 7-O-[6''-feruloyl]-glucoside/Isoscoparin 7-O-[6''-feruloyl]-glucoside/Isovitexin 7-O-[6''-p-coumaroyl]-glucoside/Apigenin 7-O-diglucoside/Apigenin 7-O-rhamnosylglucoside/Apigenin 7-O-arabinosylglucoside | | 0,40 |
| PR | PR[79] | Dehydroascorbate reductase | Defence | MT2 | MT2[114] | UPLC_UV_signal_090 | | 0,43 |
| PR | PR[79] | Dehydroascorbate reductase | Defence | MT2 | MT2[124] | UPLC_UV_signal_100 | | 0,42 |
| PR | PR[81] | Delta-1-pyrroline-5-carboxylate synthetase | Defence | MR | MR[58] | A191004-Tyramine (3TMS) | other nitrogen compounds | 0,42 |

| | | | | | | | | | |
|---|---|---|---|---|---|---|---|---|---|
| PR | PR[81] | Delta-1-pyrroline-5-carboxylate synthetase | Defence | MT2 | MT2[73] | Sinapoyl - hexose 2 | | -0,46 |
| PR | PR[101] | Fructokinase | Carbon metabolism | MT1 | MT1[104] | UPLC_UV_signal_080 | | -0,40 |
| PR | PR[152] | Malate dehydrogenase, cytoplasmic | Carbon metabolism | MR | MR[84] | A237001-NA | unknown-other | 0,38 |
| PR | PR[197] | S-adenosylmethionine synthase 2 | Nitrogen metabolism | MT1 | MT1[31] | Isoorientin 2''-O-glucoside sinapide/Isovitexin 7-O-[6''-feruloyl]-glucoside/Isoscoparin 7-O-[6''-feruloyl]-glucoside/Isovitexin 7-O-[6''-p-coumaroyl]-glucoside/Apigenin 7-O-diglucoside/Apigenin 7-O-rhamnosylglucoside/Apigenin 7-O-arabinosylglucoside | | 0,41 |
| PR | PR[197] | S-adenosylmethionine synthase 2 | Nitrogen metabolism | MT1 | MT1[128] | UPLC_UV_signal_104 | | 0,40 |
| PR | PR[198] | S-adenosylmethionine synthase 4 | Nitrogen metabolism | MT2 | MT2[31] | Isoorientin 2''-O-glucoside sinapide/Isovitexin 7-O-[6''-feruloyl]-glucoside/Isoscoparin 7-O-[6''-feruloyl]-glucoside/Isovitexin 7-O-[6''-p-coumaroyl]-glucoside/Apigenin 7-O-diglucoside/Apigenin 7-O-rhamnosylglucoside/Apigenin 7-O-arabinosylglucoside | | 0,43 |
| PR | PR[198] | S-adenosylmethionine synthase 4 | Nitrogen metabolism | MT2 | MT2[114] | UPLC_UV_signal_090 | | 0,39 |
| PR | PR[292] | Unknown protein | Unknown | ML | ML[14] | A132002-Isoleucine (2TMS) | amino acids | 0,42 |
| PR | PR[292] | Unknown protein | Unknown | MT2 | MT2[3] | 5-Feruloylquinic acid | | 0,42 |
| PR | PR[292] | Unknown protein | Unknown | MT2 | MT2[31] | Isoorientin 2''-O-glucoside sinapide/Isovitexin 7-O-[6''-feruloyl]-glucoside/Isoscoparin 7-O-[6''-feruloyl]-glucoside/Isovitexin 7-O-[6''-p-coumaroyl]-glucoside/Apigenin 7-O-diglucoside/Apigenin 7-O-rhamnosylglucoside/Apigenin 7-O-arabinosylglucoside | | 0,45 |
| PR | PR[292] | Unknown protein | Unknown | MT2 | MT2[40] | Isoscoparin 2''-O-arabinoside/Chrysoeriol 7-O-diglucoside/Isovitexin/Isoscoparin 2''-O-glucoside/Isoorientin 7-O-[6''-sinapoyl]-glucoside/Isovitexin 2''-O-rhamnoside | | 0,39 |
| PR | PR[292] | Unknown protein | Unknown | MT2 | MT2[101] | UPLC_UV_signal_077 | | 0,40 |
| PR | PR[292] | Unknown protein | Unknown | MT2 | MT2[110] | UPLC_UV_signal_086 | | 0,42 |
| PR | PR[292] | Unknown protein | Unknown | MT2 | MT2[114] | UPLC_UV_signal_090 | | 0,48 |
| PR | PR[292] | Unknown protein | Unknown | MT2 | MT2[124] | UPLC_UV_signal_100 | | 0,40 |
| ML | ML[21] | A137004-2-Piperidinecarboxylic acid (2TMS) | amino acids | MT1 | MT1[120] | UPLC_UV_signal_096 | | -0,39 |
| ML | ML[21] | A137004-2-Piperidinecarboxylic acid (2TMS) | amino acids | MT2 | MT2[20] | Feruloyl - hexose | | 0,44 |
| ML | ML[21] | A137004-2-Piperidinecarboxylic acid (2TMS) | amino acids | MT2 | MT2[21] | Feruloylquinic acid glucoside | | 0,39 |
| ML | ML[21] | A137004-2-Piperidinecarboxylic acid (2TMS) | amino acids | MT2 | MT2[34] | Isoorientin 7-O-[6''-sinapoyl]-glucoside | | 0,42 |
| ML | ML[21] | A137004-2-Piperidinecarboxylic acid (2TMS) | amino acids | MT2 | MT2[40] | Isoscoparin 2''-O-arabinoside/Chrysoeriol 7-O-diglucoside/Isovitexin/Isoscoparin 2''-O-glucoside/Isoorientin 7-O-[6''-sinapoyl]-glucoside/Isovitexin 2''-O-rhamnoside | | 0,41 |
| ML | ML[21] | A137004-2-Piperidinecarboxylic acid (2TMS) | amino acids | MT2 | MT2[53] | Isovitexin 4'-O-[6''-feruloyl]-glucoside 7-O-glucoside | | 0,45 |
| ML | ML[21] | A137004-2-Piperidinecarboxylic acid | amino acids | MT2 | MT2[66] | Isovitexin 7-O-rhamnosylglucoside/Isovitexin 4'-O- | | 0,40 |
| | | (2TMS) | | | | [6''-feruloyl]-glucoside 7-O-glucoside/Isovitexin 4'-O-[6''-sinapoyl]-glucoside 7-O-glucoside/Isoscoparin 2''-O-glucoside/Isovitexin 2''-O-glucoside | | |
| ML | ML[21] | A137004-2-Piperidinecarboxylic acid (2TMS) | amino acids | MT2 | MT2[77] | UPLC_UV_signal_053 | | 0,39 |
| ML | ML[21] | A137004-2-Piperidinecarboxylic acid (2TMS) | amino acids | MT2 | MT2[107] | UPLC_UV_signal_083 | | 0,41 |
| ML | ML[21] | A137004-2-Piperidinecarboxylic acid (2TMS) | amino acids | MT2 | MT2[110] | UPLC_UV_signal_086 | | 0,39 |
| ML | ML[21] | A137004-2-Piperidinecarboxylic acid (2TMS) | amino acids | MT2 | MT2[114] | UPLC_UV_signal_090 | | 0,44 |
| ML | ML[22] | A138001-Serine (3TMS) | amino acids | MT2 | MT2[21] | Feruloylquinic acid glucoside | | 0,40 |
| ML | ML[23] | A138002-Alanine (3TMS) | amino acids | MR | MR[95] | A300001-NA | unknown-other | 0,39 |
| ML | ML[23] | A138002-Alanine (3TMS) | amino acids | MT2 | MT2[108] | UPLC_UV_signal_084 | | 0,39 |
| ML | ML[23] | A138002-Alanine (3TMS) | amino acids | MT2 | MT2[110] | UPLC_UV_signal_086 | | 0,39 |
| ML | ML[23] | A138002-Alanine (3TMS) | amino acids | MT2 | MT2[114] | UPLC_UV_signal_090 | | 0,41 |
| ML | ML[24] | A140001-Threonine (3TMS) | amino acids | MR | MR[44] | A174008-Glutamine (4TMS) | amino acids | 0,42 |
| ML | ML[24] | A140001-Threonine (3TMS) | amino acids | MR | MR[45] | A175002-Putrescine (4TMS) | other nitrogen compounds | 0,48 |
| ML | ML[31] | A152002-Aspartic acid (3TMS) | amino acids | MR | MR[95] | A300001-NA | unknown-other | 0,39 |
| ML | ML[32] | A153002-Pyroglutamic acid (2TMS) | amino acids | MR | MR[94] | A299002-Galactinol (9TMS) | carbohydrates | -0,44 |
| ML | ML[32] | A153002-Pyroglutamic acid (2TMS) | amino acids | MR | MR[95] | A300001-NA | unknown-other | 0,48 |
| ML | ML[42] | A165001-Xylose (1MEOX) (4TMS) MP | carbohydrates | MT2 | MT2[105] | UPLC_UV_signal_081 | | 0,43 |
| ML | ML[42] | A165001-Xylose (1MEOX) (4TMS) MP | carbohydrates | MT2 | MT2[131] | UPLC_UV_signal_107 | | 0,40 |
| ML | ML[56] | A189002-Glucose (1MEOX) (5TMS) MP | carbohydrates | MT2 | MT2[78] | UPLC_UV_signal_054 | | 0,52 |
| ML | ML[56] | A189002-Glucose (1MEOX) (5TMS) MP | carbohydrates | MT2 | MT2[111] | UPLC_UV_signal_087 | | 0,43 |
| ML | ML[57] | A189007-Allantoin (4TMS) | other nitrogen compounds | MT2 | MT2[78] | UPLC_UV_signal_054 | | 0,47 |
| ML | ML[66] | A203003-NA | unknown-other | MT2 | MT2[78] | UPLC_UV_signal_054 | | 0,39 |
| ML | ML[66] | A203003-NA | unknown-other | MT2 | MT2[111] | UPLC_UV_signal_087 | | 0,39 |
| ML | ML[85] | A243003-Inositol-2-phosphate; myo- (7TMS) | carbohydrates | MT1 | MT1[3] | 5-Feruloylquinic acid | | 0,40 |
| ML | ML[94] | A299002-Galactinol (9TMS) | carbohydrates | MT2 | MT2[58] | Isovitexin 7-O-[6''-hydroxyferuloyl]-diglucoside | | -0,39 |
| ML | ML[94] | A299002-Galactinol (9TMS) | carbohydrates | MT2 | MT2[76] | UPLC_UV_signal_052 | | -0,42 |
| ML | ML[95] | A300001-NA | unknown-other | MT2 | MT2[58] | Isovitexin 7-O-[6''-hydroxyferuloyl]-diglucoside | | 0,39 |
| ML | ML[95] | A300001-NA | unknown-other | MT2 | MT2[71] | Sinapic acid derivative 1 | | 0,39 |
| ML | ML[95] | A300001-NA | unknown-other | MT2 | MT2[76] | UPLC_UV_signal_052 | | 0,45 |
| ML | ML[95] | A300001-NA | unknown-other | MT2 | MT2[82] | UPLC_UV_signal_058 | | 0,39 |
| ML | ML[95] | A300001-NA | unknown-other | MT2 | MT2[126] | UPLC_UV_signal_102 | | 0,42 |
| ML | ML[95] | A300001-NA | unknown-other | MT2 | MT2[128] | UPLC_UV_signal_104 | | 0,42 |
| ML | ML[95] | A300001-NA | unknown-other | MT2 | MT2[132] | UPLC_UV_signal_108 | | 0,40 |
| MT1 | MT1[2] | 3-p-Coumaroylquinic acid | 1 | MT2 | MT2[114] | UPLC_UV_signal_090 | | -0,39 |
| MT1 | MT1[70] | Isovitexin/Isovitexin 2''-O-rhamnoside | 1 | MT2 | MT2[67] | Isovitexin 7-O-rhamnosylglucoside/Isovitexin 4'-O- | | -0,42 |
| | | | | | | | [6''-sinapoyl]-glucoside 7-O-glucoside | |
| MT1 | MT1[73] | Sinapoyl - hexose 2 | 1 | MT2 | MT2[105] | UPLC_UV_signal_081 | | 0,42 |
| MT1 | MT1[75] | Tricin derivative | 1 | MT2 | MT2[18] | Didehydroblumenol C 2''-O-glucuronylglucoside | | 0,39 |
| MT1 | MT1[97] | UPLC_UV_signal_073 | 1 | MT2 | MT2[121] | UPLC_UV_signal_097 | | -0,40 |
| MT1 | MT1[120] | UPLC_UV_signal_096 | 1 | MT2 | MT2[3] | 5-Feruloylquinic acid | | -0,48 |
| MT1 | MT1[120] | UPLC_UV_signal_096 | 1 | MT2 | MT2[11] | Blumenol C derivative | | -0,42 |
| MT1 | MT1[120] | UPLC_UV_signal_096 | 1 | MT2 | MT2[34] | Isoorientin 7-O-[6''-sinapoyl]-glucoside | | -0,43 |
| MT1 | MT1[120] | UPLC_UV_signal_096 | 1 | MT2 | MT2[38] | Isoscoparin 2''-O-arabinoside | | -0,42 |
| MT1 | MT1[120] | UPLC_UV_signal_096 | 1 | MT2 | MT2[40] | Isoscoparin 2''-O-arabinoside/Chrysoeriol 7-O-diglucoside/Isovitexin/Isoscoparin 2''-O-glucoside/Isoorientin 7-O-[6''-sinapoyl]-glucoside/Isovitexin 2''-O-rhamnoside | | -0,43 |
| MT1 | MT1[120] | UPLC_UV_signal_096 | 1 | MT2 | MT2[43] | Isoscoparin 2''-O-glucoside/Isovitexin 2''-O-glucoside 1/Isovitexin 2''-O-arabinoside | | -0,47 |
| MT1 | MT1[120] | UPLC_UV_signal_096 | 1 | MT2 | MT2[53] | Isovitexin 4'-O-[6''-feruloyl]-glucoside 7-O-glucoside | | -0,49 |
| MT1 | MT1[120] | UPLC_UV_signal_096 | 1 | MT2 | MT2[59] | Isovitexin 7-O-[6''-hydroxyferuloyl]-glucoside 1 | | -0,40 |
| MT1 | MT1[120] | UPLC_UV_signal_096 | 1 | MT2 | MT2[66] | Isovitexin 7-O-rhamnosylglucoside/Isovitexin 4'-O-[6''-feruloyl]-glucoside 7-O-glucoside/Isovitexin 4'-O-[6''-sinapoyl]-glucoside 7-O-glucoside/Isoscoparin 2''-O-glucoside/Isovitexin 2''-O-glucoside | | -0,50 |
| MT1 | MT1[120] | UPLC_UV_signal_096 | 1 | MT2 | MT2[67] | Isovitexin 7-O-rhamnosylglucoside/Isovitexin 4'-O-[6''-sinapoyl]-glucoside 7-O-glucoside | | -0,40 |
| MT1 | MT1[120] | UPLC_UV_signal_096 | 1 | MT2 | MT2[70] | Isovitexin/Isovitexin 2''-O-rhamnoside | | -0,39 |
| MT1 | MT1[120] | UPLC_UV_signal_096 | 1 | MT2 | MT2[90] | UPLC_UV_signal_066 | | -0,40 |
| MT1 | MT1[120] | UPLC_UV_signal_096 | 1 | MT2 | MT2[108] | UPLC_UV_signal_084 | | -0,43 |
| MT1 | MT1[120] | UPLC_UV_signal_096 | 1 | MT2 | MT2[110] | UPLC_UV_signal_086 | | -0,46 |
| MT1 | MT1[120] | UPLC_UV_signal_096 | 1 | MT2 | MT2[114] | UPLC_UV_signal_090 | | -0,45 |
| MT1 | MT1[120] | UPLC_UV_signal_096 | 1 | MT2 | MT2[118] | UPLC_UV_signal_094 | | -0,46 |
| MT1 | MT1[120] | UPLC_UV_signal_096 | 1 | MT2 | MT2[124] | UPLC_UV_signal_100 | | -0,44 |
| MT1 | MT1[120] | UPLC_UV_signal_096 | 1 | MT2 | MT2[126] | UPLC_UV_signal_102 | | -0,41 |
| MT1 | MT1[131] | UPLC_UV_signal_107 | 1 | MT2 | MT2[14] | Caffeoyl - hexose | | 0,41 |
| MT1 | MT1[131] | UPLC_UV_signal_107 | 1 | MT2 | MT2[130] | UPLC_UV_signal_106 | | 0,42 |
| MT1 | MT1[132] | UPLC_UV_signal_108 | 1 | MT2 | MT2[18] | Didehydroblumenol C 2''-O-glucuronylglucoside | | 0,39 |
| MT1 | MT1[132] | UPLC_UV_signal_108 | 1 | MT2 | MT2[130] | UPLC_UV_signal_106 | | 0,40 |
| MT1 | MT1[134] | UPLC_UV_signal_110 | 1 | MT2 | MT2[12] | Blumenol C glucoside | | 0,42 |
| MT1 | MT1[134] | UPLC_UV_signal_110 | 1 | MT2 | MT2[14] | Caffeoyl - hexose | | 0,49 |
| MT1 | MT1[134] | UPLC_UV_signal_110 | 1 | MT2 | MT2[31] | Isoorientin 2''-O-glucoside sinapide/Isovitexin 7-O-[6''-feruloyl]-glucoside/Isoscoparin 7-O-[6''-feruloyl]-glucoside/Isovitexin 7-O-[6''-p-coumaroyl]-glucoside/Apigenin 7-O-diglucoside/Apigenin 7-O-rhamnosylglucoside/Apigenin 7-O-arabinosylglucoside | | 0,41 |
| MT1 | MT1[134] | UPLC_UV_signal_110 | 1 | MT2 | MT2[108] | UPLC_UV_signal_084 | | 0,46 |
| MT1 | MT1[134] | UPLC_UV_signal_110 | 1 | MT2 | MT2[118] | UPLC_UV_signal_094 | | 0,52 |
| MT1 | MT1[134] | UPLC_UV_signal_110 | 1 | MT2 | MT2[124] | UPLC_UV_signal_100 | | 0,58 |
| MT1 | MT1[134] | UPLC_UV_signal_110 | 1 | MT2 | MT2[130] | UPLC_UV_signal_106 | | 0,43 |

Table S2. Statistically significant correlations between phenotypic traits and omic traits (corresponding to edges in Figure 4)

| Group of traits | Trait ID | Trait name | Category | Group of traits | Trait ID | Trait name | Category | Correlation |
|---|---|---|---|---|---|---|---|---|
| PHG | PHG[1] | 1000-grain weight | 1 | PL | PL[191] | Unknown protein | Unknown | -0,38 |
| PHG | PHG[1] | 1000-grain weight | 1 | ML | ML[25] | A140005-Threonic acid-1;4-lactone (2TMS) | sugar acids | 0,35 |
| PHG | PHG[1] | 1000-grain weight | 1 | MT2 | MT2[78] | UPLC_UV_signal_054 | 1 | -0,36 |
| PHG | PHG[1] | 1000-grain weight | 1 | MT2 | MT2[84] | UPLC_UV_signal_060 | 1 | -0,35 |
| PHG | PHG[1] | 1000-grain weight | 1 | MT2 | MT2[121] | UPLC_UV_signal_097 | 1 | -0,37 |
| PHG | PHG[3] | Length of main stem | 1 | PL | PL[13] | Ascorbate peroxidase | Defence | -0,35 |
| PHG | PHG[3] | Length of main stem | 1 | PL | PL[126] | Ribulose bisphosphate carboxylase/oxygenase activase A | Photosynthesis | -0,51 |
| PHG | PHG[3] | Length of main stem | 1 | PL | PL[128] | Ribulose bisphosphate carboxylase/oxygenase activase A | Photosynthesis | -0,38 |
| PHG | PHG[3] | Length of main stem | 1 | PL | PL[201] | Unknown protein | Unknown | -0,36 |
| PHG | PHG[3] | Length of main stem | 1 | ML | ML[24] | A140001-Threonine (3TMS) | amino acids | 0,34 |
| PHG | PHG[3] | Length of main stem | 1 | ML | ML[53] | A185001-Quinic acid (5TMS) | other carboxylic acids | -0,40 |
| PHG | PHG[3] | Length of main stem | 1 | ML | ML[54] | A185002-Dehydroascorbic acid dimer (2MEOX) MP | sugar acids | 0,47 |
| PHG | PHG[3] | Length of main stem | 1 | ML | ML[55] | A187002-Fructose (1MEOX) (5TMS) MP | carbohydrates | -0,51 |
| PHG | PHG[3] | Length of main stem | 1 | ML | ML[56] | A189002-Glucose (1MEOX) (5TMS) MP | carbohydrates | -0,35 |
| PHG | PHG[3] | Length of main stem | 1 | ML | ML[58] | A191004-Tyramine (3TMS) | other nitrogen compounds | 0,36 |
| PHG | PHG[3] | Length of main stem | 1 | ML | ML[66] | A203003-NA | unknown-other | -0,36 |
| PHG | PHG[3] | Length of main stem | 1 | ML | ML[83] | A236005-NA | unknown-other | 0,48 |
| PHG | PHG[3] | Length of main stem | 1 | ML | ML[94] | A299002-Galactinol (9TMS) | carbohydrates | 0,38 |
| PHG | PHG[3] | Length of main stem | 1 | ML | ML[95] | A300001-NA | unknown-other | -0,36 |
| PHG | PHG[4] | Straw weight per plant | 1 | PR | PR[117] | Glutathione S-transferase | Defence | 0,33 |
| PHG | PHG[4] | Straw weight per plant | 1 | ML | ML[55] | A187002-Fructose (1MEOX) (5TMS) MP | carbohydrates | -0,39 |
| PHG | PHG[4] | Straw weight per plant | 1 | MR | MR[6] | A119001-Phosphoric acid monomethyl ester (2TMS) | unknown-other | 0,34 |
| PHG | PHG[4] | Straw weight per plant | 1 | MR | MR[44] | A174008-Glutamine (4TMS) | amino acids | 0,37 |
| PHG | PHG[4] | Straw weight per plant | 1 | MT1 | MT1[20] | Feruloyl - hexose | 1 | 0,34 |
| PHG | PHG[7] | Grain weight per main spike | 1 | ML | ML[55] | A187002-Fructose (1MEOX) (5TMS) MP | carbohydrates | -0,37 |
| PHG | PHG[9] | Number of spikelets per main spike | 1 | ML | ML[55] | A187002-Fructose (1MEOX) (5TMS) MP | carbohydrates | -0,33 |
| PHG | PHG[9] | Number of spikelets per main spike | 1 | ML | ML[56] | A189002-Glucose (1MEOX) (5TMS) MP | carbohydrates | -0,35 |
| PHG | PHG[9] | Number of spikelets per main spike | 1 | MT2 | MT2[78] | UPLC_UV_signal_054 | 1 | -0,34 |
| PHG | PHG[10] | Length of main spike | 1 | PL | PL[126] | Ribulose bisphosphate carboxylase/oxygenase activase A | Photosynthesis | -0,44 |
| PHG | PHG[10] | Length of main spike | 1 | ML | ML[3] | A110002-Hydroxylamine (3TMS) | other nitrogen compounds | 0,33 |
| PHG | PHG[10] | Length of main spike | 1 | ML | ML[24] | A140001-Threonine (3TMS) | amino acids | 0,35 |
| PHG | PHG[10] | Length of main spike | 1 | ML | ML[53] | A185001-Quinic acid (5TMS) | other carboxylic acids | -0,34 |
| PHG | PHG[10] | Length of main spike | 1 | ML | ML[54] | A185002-Dehydroascorbic acid dimer (2MEOX) MP | sugar acids | 0,39 |
| PHG | PHG[10] | Length of main spike | 1 | ML | ML[55] | A187002-Fructose (1MEOX) (5TMS) MP | carbohydrates | -0,58 |
| PHG | PHG[10] | Length of main spike | 1 | ML | ML[56] | A189002-Glucose (1MEOX) (5TMS) MP | carbohydrates | -0,39 |
| PHG | PHG[10] | Length of main spike | 1 | ML | ML[58] | A191004-Tyramine (3TMS) | other nitrogen compounds | 0,40 |
| PHG | PHG[10] | Length of main spike | 1 | ML | ML[66] | A203003-NA | unknown-other | -0,33 |
| PHG | PHG[10] | Length of main spike | 1 | ML | ML[83] | A236005-NA | unknown-other | 0,46 |
| PHG | PHG[10] | Length of main spike | 1 | ML | ML[94] | A299002-Galactinol (9TMS) | carbohydrates | 0,35 |
| PHG | PHG[10] | Length of main spike | 1 | ML | ML[95] | A300001-NA | unknown-other | -0,37 |
| PHG | PHG[10] | Length of main spike | 1 | MR | MR[44] | A174008-Glutamine (4TMS) | amino acids | 0,40 |
| PHG | PHG[11] | Grain weight per lateral spike | 1 | PR | PR[208] | Transketolase | Carbon metabolism | -0,37 |
| PHG | PHG[14] | Length of lateral spike | 1 | MR | MR[44] | A174008-Glutamine (4TMS) | amino acids | 0,34 |
| PHG | PHG[14] | Length of lateral spike | 1 | MR | MR[45] | A175002-Putrescine (4TMS) | other nitrogen compounds | 0,39 |
| PHG | PHG[18] | Heading stage | 1 | MT1 | MT1[50] | Isovitexin 2''-O-glucoside derivative | 1 | -0,36 |
| PHG | PHG[18] | Heading stage | 1 | MT1 | MT1[71] | Sinapic acid derivative 1 | 1 | -0,36 |
| PHG | PHG[18] | Heading stage | 1 | MT2 | MT2[114] | UPLC_UV_signal_090 | 1 | 0,34 |
| PHG | PHG[20] | Fertility of main spike | 1 | MR | MR[44] | A174008-Glutamine (4TMS) | amino acids | -0,36 |
| PHG | PHG[20] | Fertility of main spike | 1 | MR | MR[45] | A175002-Putrescine (4TMS) | other nitrogen compounds | -0,35 |
| PHG | PHG[20] | Fertility of main spike | 1 | MR | MR[52] | A182004-Citric acid (4TMS) | Krebs cycle acids | -0,38 |
| PHG | PHG[21] | Fertility of lateral spike | 1 | PR | PR[197] | S-adenosylmethionine synthase 2 | Nitrogen metabolism | -0,34 |
| PHG | PHG[22] | Protein content in grain | 1 | ML | ML[54] | A185002-Dehydroascorbic acid dimer (2MEOX) MP | sugar acids | 0,36 |
| PHG | PHG[24] | Straw weight per pot | 1 | PR | PR[57] | Betaine aldehyde dehydrogenase | Defence | 0,33 |
| PHG | PHG[24] | Straw weight per pot | 1 | ML | ML[53] | A185001-Quinic acid (5TMS) | other carboxylic acids | -0,37 |
| PHG | PHG[24] | Straw weight per pot | 1 | MR | MR[26] | A144001-Alanine; beta- (3TMS) | amino acids | 0,35 |
| PHG | PHG[24] | Straw weight per pot | 1 | MT1 | MT1[84] | UPLC_UV_signal_060 | 1 | -0,40 |
| PHG | PHG[25] | Harvest index | 1 | MR | MR[44] | A174008-Glutamine (4TMS) | amino acids | -0,35 |
| PHG | PHG[25] | Harvest index | 1 | MR | MR[46] | A177001-Ribonic acid (5TMS) | sugar acids | -0,36 |
| PHG | PHG[25] | Harvest index | 1 | MR | MR[89] | A264001-Sucrose (8TMS) | carbohydrates | -0,35 |